# Impact of the crystal orientation on spin-orbit torques in Fe/Pd bilayers

Ankit Kumar[1], Nilamani Behera[1], Rahul Gupta[1], Sajid Husain[1], Henry Stopfel[1], Vassilios Kapaklis[2], Rimantas Brucas[1], and Peter Svedlindh[1]

[1]. Department of Engineering Sciences, Uppsala University, Box 534, SE-751 21 Uppsala, Sweden

[2]. Department of Physics and Astronomy, Uppsala University, Box 516, SE-751 20 Uppsala, Sweden

Spin-orbit torques in ferromagnetic (FM)/non-magnetic (NM) heterostructures offer more energy-efficient means to realize spin-logic devices; however, their strengths are determined by the heterostructure interface. This work examines crystal orientation impact on the spin-orbit torque efficiency in different Fe/Pd bilayer systems. Spin torque ferromagnetic measurements evidence that the damping-like torque efficiency is higher in epitaxial than in polycrystalline bilayer structures while the field-like torque is negligible in all bilayer structures. The strength of the damping-like torque decreases with deterioration of the bilayer epitaxial quality. The present finding provides fresh insight for the enhancement of spin-orbit torques in magnetic heterostructures.

Spin torque based spin-logic devices have been identified as prime candidates for beyond Moore technologies because of their good scaling and significant potential to operate at ultra-low power and high speed. Spin torques can be classified as spin transfer torques (STTs) and spin-orbit torques (SOTs) [1, 2]. STT based spin logics have limitations due to the stochastic nature of the STT-switching and a high switching critical current density, which makes these devices unsuitable for ultrafast operation at the sub-nanosecond regime and beyond. However, SOT based logic operation relying on torques generated by the net spin accumulation at the magnetic/nonmagnetic interface has the potential to overcome these limitations. A basic building block for spin-orbit torque devices is the ferromagnetic (FM) and nonmagnetic (NM) bilayer with its FM/NM interface, as in SOT-MRAMs and spin oscillators. The net spin accumulation at the interface can be generated by several methods, e.g. through the Edelstein effect [3], spin Hall effect (SHE) [4], Seebeck effect [5], etc. An elegant way to generate a SOT is provided by the SHE, which may exhibit both bulk and interfacial contributions [4]. The SHE generated spin current exerts a torque on the magnetization of the free magnetic layer via a spin angular momentum transfer mechanism between the different orbitals; known as SOT. This SOT can be damping- or field-like depending on the FM/NM interface [2, 4]. Zhang *et al.* [6] have reported the importance of the interfacial transparency for the strength of the SOT in polycrystalline FM/Pt (FM = Py, Co, $Co_{1-x}Ni_x$) heterostructures. They concluded that the electronic band matching is important for the enhancement of the SOT in bilayers. Later, Zhou *et al.* [7] reported the effect of a collinear antiferromagnetic state on the SOT efficiency in $Py/L1_0$-IrMn bilayers. It was concluded that the coherent collinear magnetic order at the interface induces a large SOTs efficiency. Lee *et al.* [8] have reported significant enhancement of the SOT in the polycrystalline Pt/CoFeB system by the interface modification involving a Ti layer, which was attributed to enhancement of the transparency and reduction of the magnetic proximity at the interface. Furthermore, it has also been evidenced that the dominant spin relaxation or SHE mechanism in metallic NM layers follows the Elliot–Yafet mechanism, where the spin Hall angle is proportional to its resistivity [2, 4, 9, 10]. However, it is also reported that in epitaxial NM Ta and Pt layers, the D'yakonov–Perel' mechanism originating from interfacial spin-orbit interaction (SOI) provides the dominating spin relaxation mechanism [11, 12, 13]. In case of the D'yakonov–Perel' mechanism, spin Hall angle is independent of the NM layer resistivity. Therefore, it is intriguing to understand the SHE and associated SOT efficiency in epitaxial magnetic heterostructures that are devoid of or exhibit very small interfacial Rashba SOI.

The interfacial spin torques are very sensitive to the crystallographic structure and therefore to the orbital ordering at the interface. Hence, in this study we have examined the effect of crystallographically ordered and disordered interfaces on the SOT efficiency in the Fe/Pd system. We have performed spin torque ferromagnetic resonance (ST-FMR) measurements at different applied dc currents to extract the change of the effective damping of the Fe/Pd bilayer in three different epitaxial heterostructures and one polycrystalline heterostructure. The observed critical current density at which the effective damping reverses its sign is significantly lower in the epitaxial Fe/Pd structures than in the polycrystalline Fe/Pd structure, which is evidence of a dominating of anti-damping spin torque in the epitaxial Fe/Pd structures.

Fe/Pd bilayers were deposited on Si/SiO$_2$ and MgO substrates using an ultra-high vacuum sputtering system. To remove surface contamination and to improve surface quality of the substrates, the substrates were heat treated at 620 °C for 2 hours prior to deposition. The polycrystalline Fe/Pd bilayer was grown at room temperature on a Si(100)/SiO$_2$ substrate, while the different epitaxial Fe/Pd bilayers were grown on MgO(100) substrates at substrate temperatures of 238 °C, 288 °C and 338 °C. The thickness of the Fe and Pd layers were kept constant of 5 nm, respectively. The polycrystalline film is hereafter referred to as poly-Fe/Pd, while the epitaxial Fe/Pd films are named as epi-Fe/Pd_238, epi-Fe/Pd_288 and epi-Fe/Pd_338, respectively.

The crystallographic orientation and mosaicity of the epitaxial bilayers were examined by performing X-ray diffraction (XRD) measurements. The individual layer thickness and interface roughness were obtained using X-ray reflectivity (XRR) measurements. The scans covered the 2θ range 0°– 6°, and the XRR results were analysed using the PANaltical X'pert Reflectivity software package with a combined genetic and segmented algorithm model.

SOT measurements were performed by employing ST-FMR technique [14]. ST-FMR spectra were recorded on 10 μm (width, $w$) × 100 μm (length, $l$) patterned Fe(5nm)/Pd(5nm) structures along magnetization easy axis. The measured resistances of poly-Fe/Pd, epi-Fe/Pd_238, epi-Fe/Pd_288, and epi-Fe/Pd_338 patterned structures are $R_{\text{poly}-\text{Fe/Pd}} = 179 \, \Omega$, $R_{\text{epi}-\text{Fe/Pd}\_238} = 117 \, \Omega$, $R_{\text{epi}-\text{Fe/Pd}\_288} = 115 \, \Omega$ and $R_{\text{epi}-\text{Fe/Pd}\_338} = 109 \, \Omega$, respectively. In ST-FMR measurements, the microwave current ($I_{rf}$) was injected along the sample length; see Fig. 1(a) of Ref. [14]. The ST-FMR spectra were recorded by scanning the in-plane magnetic field at 45° ($\varphi$) with respect to the direction of $I_{rf}$ at different constant frequencies ranging from 9 to 16 GHz and for different applied dc current density ($J_c$). These measurements used an internal amplitude modulation technique, where a 50% amplitude modulation of the microwave signal at 211 Hz was used for lock-in detection. The applied microwave power was kept constant at 10 dBm during measurements (for more details of the experimental setup, see Ref. [14]). The non-uniformity of the microwave power inside the patterned bar is negligible for the studied structures.

Figure 1 (a) shows the XRD $\theta - 2\theta$ pattern of the epi-Fe/Pd_238 film, which clearly evidence the Fe(200) and Pd(200) crystallographic orientations. It is known that the Fe(100) crystallographic plane rotates 45° in-plane to match the plane of MgO(100) while the Pd(100) plane rotates 45° with respect to Fe(100) to grow epitaxially on the Fe(100) film. To confirm the epitaxial growth, $\omega$ scans were performed across the Fe(200) and Pd(200) orientations as shown in Figs. 1(b) and 1(c), respectively. The presence of sharp single peak in respective rocking curve ($\omega$ scan) confirm the epitaxial orientation of the Fe(200)/Pd(200) bilayers. The $\omega$ scan determined value of the full width at half maximum (FWHM) for the Fe(200) orientation for the epi-Fe/Pd_238, epi-Fe/Pd_288 and epi-Fe/Pd_338 films are 2.8°, 3.0°, and 3.5°, respectively, while the FWHM (0.15°) is nearly constant for Pd(200) peak for all films. Figure 2 shows XRR patterns along with simulated lines for bilayer films. The XRR determined values for individual layer thicknesses and roughness are presented in Table 1. The interface roughness of polycrystalline and epi-Fe/Pd bilayers is in the range of 0.39 – 0.67 nm, confirming that the different bilayer samples exhibit sharp interfaces. There is no significant difference in interface roughness comparing polycrystalline and epi-Fe/Pd bilayers.

Figure 3(a) shows a schematic of the spin-orbit torques acting on the Fe layer applying a constant dc current density $J_c$ in the Pd layer. The applied $J_c$ creates a spin accumulation at the Fe/Pd interface that acts as damping-like (DL) and field-like (FL) torques on the Fe layer magnetization $M$ [15, 16]. In ST-FMR measurements, the microwave current $I_{rf}$ in the NM layer generates an Oersted field ($H_{rf}$) and a transverse spin current density ($J_s$) via the SHE [2, 4, 17]. Conversely, in a case of interfacial SOI and symmetry breaking at the FM/NM interface, an Rashba Edelstein effective field is generated whose direction will be opposite to that of the Oersted field [2, 4]. The $I_{rf}$ excited temporal variation of the magnetization vector induces a time varying resistance due to anisotropic magnetoresistance of the FM layer. The varying resistance mixing with $I_{rf}$ yields a dc voltage output. At resonance, the torques due to the Oersted field and the transverse spin current contribute with anti-symmetric and symmetric profiles, respectively, to the FMR line-shape, while the Rashba Edelstein effect generated field-like torque contributes with an anti-symmetric profile [2-4, 15-17].

The SOT induced time evolution of the magnetization vector $\vec{m} = \vec{M}/M_s$ is governed by the Landau-Lifshitz-Gilbert (LLG) equation of motion, expressed as [6, 18]

$$\frac{\partial \hat{m}}{\partial t} = \frac{-\gamma}{(1+\alpha^2)}\left(\hat{m} \times \mu_0 \vec{H}_{eff}\right) - \frac{\gamma\alpha}{(1+\alpha^2)}\left(\hat{m} \times \hat{m} \times \mu_0 \vec{H}_{eff}\right) + \\ + \frac{\gamma \hbar J_S}{(1+\alpha^2) 2eM_s t_{Fe}}(\hat{m} \times \hat{\sigma} \times \hat{m}) - \frac{\gamma}{(1+\alpha^2)}\left(\hat{m} \times \mu_0 \vec{H}_t\right), \quad (1)$$

where $\gamma$ is the gyro magnetic ratio, $\alpha$ is the (Gilbert) damping parameter, $M_s$ is the saturation magnetization, $\vec{\sigma}$ is the direction of the injected spin moment, and $t_{Fe}$ the thickness of the Fe layer on which the spin torque is acting, and $\vec{H}_t = \vec{H}_{rf} + \vec{H}_{FL}$; $\vec{H}_{rf}$ and $\vec{H}_{FL}$ are the Oersted field and Rashba Edelstein field contributions, respectively. The SOTs acting on the Fe layer magnetization are DL torque due to the transverse spin current; $\frac{\gamma \hbar J_S}{(1+\alpha^2) 2eM_s t_{Fe}}(\hat{m} \times \hat{\sigma} \times \hat{m})$, and the field-like torque due to the Oersted and Rashba Edelstein fields; $\frac{\gamma}{(1+\alpha^2)}\left(\hat{m} \times \mu_0 \vec{H}_t\right)$.

The recorded ST-FMR spectra of the poly-Fe/Pd and epi-Fe/Pd_238 bilayers at 11 GHz is shown in Fig. 3(b). The recorded voltage signal, which exhibits symmetric and anti-symmetric Lorentzian weight factors, is larger in epitaxial Fe/Pd structures compared to the polycrystalline Fe/Pd structure. The ST-FMR spectrum is fitted using expression [6, 17]

$$V_{mix} = V_S \frac{\left(\frac{\Delta H}{2}\right)^2}{\left(\frac{\Delta H}{2}\right)^2 + (H-H_r)^2} + V_A \frac{\frac{\Delta H}{2}(H-H_r)}{\left(\frac{\Delta H}{2}\right)^2 + (H-H_r)^2}, \quad (2)$$

where $V_S$ and $V_A$ are the amplitudes of the symmetric and anti-symmetric components, respectively, of $V_{mix}$. $\Delta H$ and $H_r$ are the FWHM linewidth and resonance field, respectively. $V_S$ is proportional to the *in-plane* DL effective torque and $V_A$ is proportional to the *out-of-plane* effective torque due to the Oersted field and Rashba Edelstein field-like torques; for more details see Refs. [16-18]. The fitting determined values of $\mu_0 \Delta H$ vs. frequency ($f$) and $f$ vs. $H_r$ for the poly-Fe/Pd and epi-Fe/Pd_238 structures at different applied dc currents $I_{dc}$ are presented in Figs. 3(c) and 3(d), and in Figs. 3(e) and 3(f), respectively. Absence of an $I_{dc}$-dependence of the $f$ vs. $H_r$ profiles implies negligible field-like torque in all studied Fe/Pd bilayers. The $\mu_0 \Delta H$ vs. $f$ data at different applied $I_{dc}$ were fitted to determine the effective damping $\alpha_{eff}$ using the expression $\mu_0 \Delta H = \frac{4\pi \alpha_{eff}}{\gamma} f + \mu_0 \Delta H_0$, where $\Delta H_0$ is the frequency independent contribution and $\gamma/2\pi \approx 28.03$ GHz/T is gyromagnetic ratio. The $f$ vs. $\mu_0 H_r$ data were fitted using the in-plane Kittel equation, yielding $\mu_0 M_{eff} \approx 1.70(1)$ $T$ and $\approx 2.05(1)$ $T$ for poly_Fe/Pd and epi_Fe/Pd bilayers, respectively.

The $\alpha_{eff}(I_{dc})$ values determined for positive and negative field scans are presented in Fig. 4 for all Fe/Pd structures.

The $I_{dc}$ dependent changes of the effective Gilbert damping $\alpha_{eff}(I_{dc})$ is expressed as $\alpha_{eff}(I_{dc}) - \alpha_{eff}(I_{dc} = 0) = \left(\frac{\sin\varphi}{(H_r + 0.5 M_{eff})\mu_0 M_S t_{Fe}} \frac{\hbar}{2e}\right) J_c \theta_{cs}$, where $J_c = \frac{I_{dc}}{A_{Pd}} \frac{R_{Fe/Pd}}{R_{Pd}}$ is the current density in the Pd layer, $A_{Pd}$ ($= 4.5 \times 10^{-14}$ m$^2$) is the Pd layer cross sectional area, and $\theta_{cs}$ is the effective charge-to-spin conversion efficiency [19]. In all of the Fe/Pd layers 50% of the applied dc current passes through the Pd layer. The $J_c$ dependent percentage change of the effective damping for poly-Fe/Pd, epi-Fe/Pd_238, epi-Fe/Pd_288 and epi-Fe/Pd_338 are 1.95%, 7.05%, 6.90% and 6.69 %, respectively, at $J_c = \pm 3 \times 10^{11}$ $\frac{A}{m^2}$. The critical current density ($J_{c,crit}$) or damping-like torque efficiency at which the effective damping becomes zero or reverses sign in respective Fe/Pd structures is determined by using the applied dc current dependent changes of the effective damping values. The slope determined $J_{c,crit}$ values are $\pm 1.54 \times 10^{13}$ $\frac{A}{m^2}$, $\pm 4.25 \times 10^{12}$ $\frac{A}{m^2}$, $\pm 4.34 \times 10^{12}$ $\frac{A}{m^2}$, and $\pm 4.48 \times 10^{12}$ $\frac{A}{m^2}$ for poly-Fe/Pd, epi-Fe/Pd_238, epi-Fe/Pd_288, epi-Fe/Pd_338, respectively.

The $\alpha_{eff}(I_{dc} = 0)$ value for poly-Fe/Pd is higher than the values for epi-Fe/Pd bilayers, while the $J_c$ dependent percentage change of the effective damping is lower in poly-Fe/Pd. The damping-like torque strength decreases for the epitaxial Fe/Pd bilayers grown at higher temperatures, which appears to linked with the increase of mosaicity in the high temperatures grown films. These results infer an impact of the crystallographic orientation on the $J_c$ dependent effective damping modulation.

Very recently, spin-orbit torques have been studied in perpendicularly magnetized epitaxial and polycrystalline Co/Pt bilayers by using first and second harmonic Hall resistance measurements [20]. In that study, the damping-like torque in epitaxial Co/Pt was found to be 1.3 times smaller than the torque in polycrystalline Co/Pt. These results are in contrast to our results for *in-plane* magnetized Fe/Pd bilayers. This discrepancy results from the large spin-orbit coupling (SOC) of the Pt layer (5d group element) and that for the epitaxial bilayer one expects an abrupt potential change at the epitaxial interface, and therefore a strong interface spin loss [21]. In our study the Pd layers (4d group element) exhibit a low SOC strength. According to the XRR results there is no significant difference in interface roughness between the Fe/Pd bilayers. Therefore, one expects an abrupt potential change in all Fe/Pd bilayers; hence, interface spin loss cannot explain the different $\alpha_{eff}(I_{dc})$ results among the Fe/Pd bilayers.

The Elliot–Yafet mechanism provides the dominant contribution to the SHE in inversion symmetry preserved metals and it can be modulated by crystal structure and crystallographic orientation [22]. However, for Pd being a cubic system the SHE is expected to exhibit weak crystallographic orientation dependence [22]. The Pd layer thickness is less than its mean free path (8 nm) [21], therefore the effect of the epitaxial interface on the damping-like torque efficiency cannot be ignored. Further, the D'yakonov–Perel' mechanism is observed in epitaxial 5d Ta and Pt films, which exhibit strong interfacial SOI. In our studied Fe/Pd heterstructures we have not observed any interfacial Rashba coupling, which indicates negligible interfacial SOI. The D'yakonov–Perel' mechanism is also proposed to describe the interface in metal heterostructures [23], but detailed theoretical and experimental understanding require to unveil the spin torques mechanism and the spin relaxation phenomena at epitaxial metallic interfaces with weak interfacial SOI, alike in Fe/Pd bilayers.

In conclusion, we have examined the impact of the crystallographic orientation in Fe/Pd bilayer structures on the spin angular momentum transfer across the interface by performing ST-FMR measurements. The effective damping value of the polycrystalline Fe/Pd bilayer is approximately two times larger than the values of the epitaxial Fe/Pd bilayers. The studied Fe/Pd bilayers exhibit only damping-like spin-orbit toque and its strength is significantly higher in the epitaxial heterostructures as compared to the polycrystalline structure. These results provide directions for the realization of energy efficient spin-logic operations, or more specifically nano-oscillators for neuromorphic computing by utilization epitaxial magnetic heterostructures.

ACKNOWLEDGEMENT

This work is supported by the Swedish Research Council (VR), grant no 2017-03799.

Figure 1. (a) X-ray diffraction θ-2θ patterns of poly-Fe/Pd and epi-Fe/Pd_238. Rocking curve scans at (b) Fe(200) peak and (c) Pd(200) peak for the different epi-Fe/Pd bilayers.

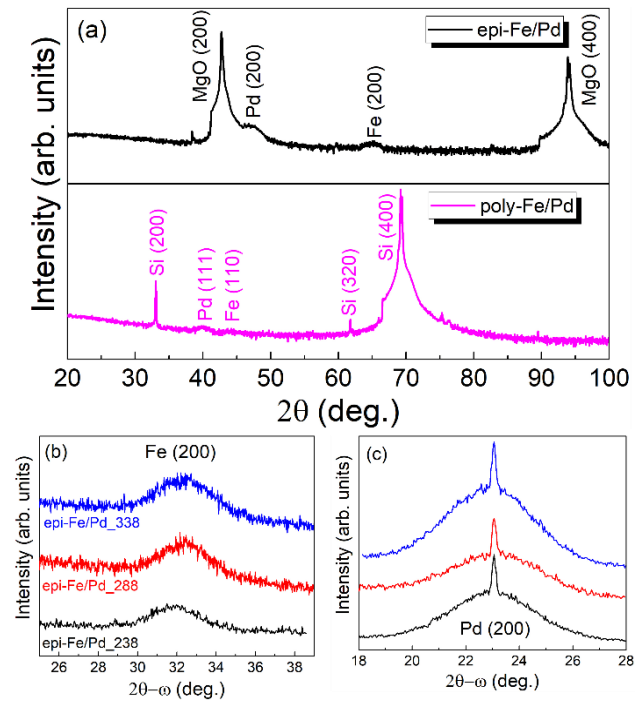

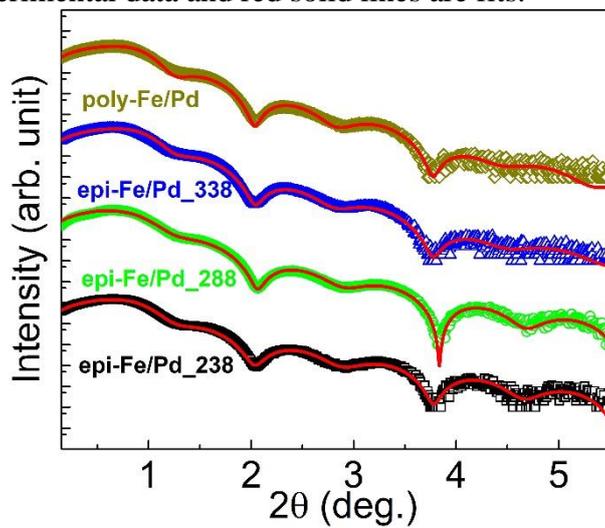

Fig 2. XRR spectra of the poly-Fe/Pd, epi-Fe/Pd_238, epi-Fe/Pd_288 and epi-Fe/Pd_338 films. Symbols are experimental data and red solid lines are fits.

Figure 3. (a) Schematic of the spin-orbit torques acting in the Fe/Pd bilayer. (b) Comparison of ST-FMR spectra of the poly-Fe/Pd and epi-Fe/Pd_232 structures at 10 GHz. $\mu_0 \Delta H$ vs. $f$ of (c) poly-Fe/Pd and (d) epi-Fe/Pd_238. $f$ vs. $H_r$ of (e) poly-Fe/Pd and (f) epi-Fe/Pd_238.

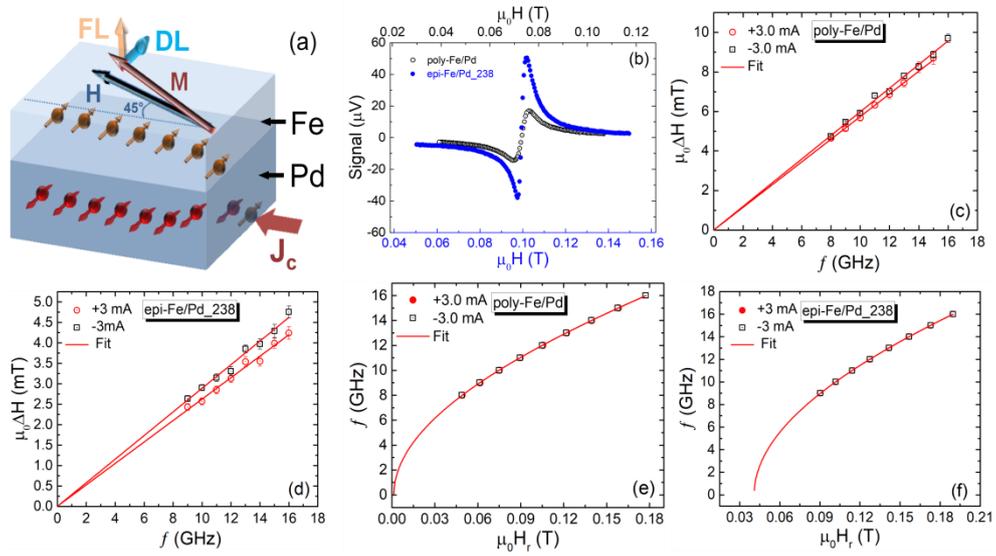

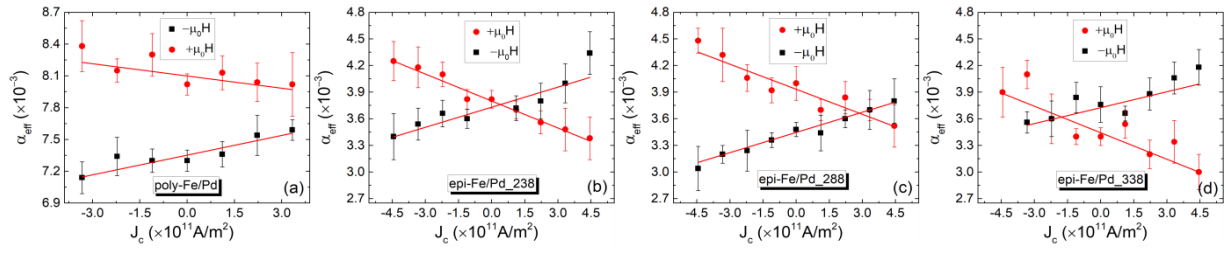

Figure 4. Applied dc current density dependent changes of the effective damping in (a) poly-Fe/Pd, (b) epi-Fe/Pd_238, (c) epi-Fe/Pd_288 and (d) epi-Fe/Pd_338 structures.

Table:1.

| Sample | $t_{Fe}$ (nm) (±0.02) | $t_{Pd}$ (nm) (±0.02) | $\sigma(nm)$ (±0.03) |
|---|---|---|---|
| epi-Fe/Pd_238 | 4.94 | 4.64 | 0.39 |
| epi-Fe/Pd_288 | 4.96 | 4.56 | 0.56 |
| epi-Fe/Pd_338 | 5.20 | 4.61 | 0.65 |
| poly-Fe/Pd | 5.15 | 4.47 | 0.44 |

Table1: XRR fitting parameters; thickness and roughness/interface width in the poly-Fe/Pd, epi-Fe/Pd_238, epi-Fe/Pd_288 and epi-Fe/Pd_338 bilayers. Here $t_{Fe}$, $t_{Fe}$, and $\sigma$ refer to the thickness of Fe layer, thickness of Pd layer and roughness of the interface, respectively.